\documentclass{aa}  

\usepackage[hang, flushmargin]{footmisc}

\usepackage{graphicx}
\usepackage{txfonts}
\usepackage{tikz,xcolor,hyperref}

\hypersetup{
        colorlinks = true,
        urlcolor = blue,
        linkcolor = blue,
        citecolor=blue
}

\definecolor{lime}{HTML}{A6CE39}
\DeclareRobustCommand{\orcidicon}{%
        \begin{tikzpicture}
        \draw[lime, fill=lime] (0,0) 
        circle [radius=0.16] 
        node[white] {{\fontfamily{qag}\selectfont \tiny ID}};
        \draw[white, fill=white] (-0.0625,0.095) 
        circle [radius=0.007];
        \end{tikzpicture}
        \hspace{-2mm}
}

\foreach \x in {A, ..., Z}{%
\expandafter\xdef\csname orcid\x\endcsname{\noexpand\href{https://orcid.org/\csname orcidauthor\x\endcsname}{\noexpand\orcidicon}}
}

\begin{document}

    \title{Parametric study of the kinematic evolution of coronal mass ejection shock waves and their relation to flaring activity}
    
    \titlerunning{Kinematic properties of CME / coronal shocks}

    \author{Manon Jarry\orcidA{}
    \inst{1}
    \and
    Alexis P. Rouillard\orcidB{}
    \inst{1}
    \and
    Illya Plotnikov\orcidC{}
    \inst{1}
    \and
    Athanasios Kouloumvakos\orcidD{}
    \inst{2}
    \and
    Alexander Warmuth\orcidE{}
    \inst{3}
    }

   \institute{IRAP, CNRS, Université Toulouse III–Paul Sabatier, Toulouse, France
   \and
   Applied Physics Laboratory, The Johns Hopkins University, Laurel, MD, United States
   \and
   Leibniz-Institut für Astrophysik Potsdam (AIP), Potsdam, Germany
   }
             
    \date{Received November 16, 2022; accepted February 17, 2023}


\abstract
    {Coronal and interplanetary shock waves produced by coronal mass ejections (CMEs) are major drivers of space-weather phenomena, inducing major changes in the heliospheric radiation environment and directly perturbing the near-Earth environment, including its magnetosphere. A better understanding of how these shock waves evolve from the corona to the interplanetary medium can therefore contribute  to improving nowcasting and forecasting of  space weather. Early warnings from these shock waves can come from radio measurements as well as coronagraphic observations that can be exploited to characterise the dynamical evolution of these structures.}
    {Our aim is to analyse the geometrical and kinematic properties of 32 CME shock waves derived from multi-point white-light and ultraviolet imagery taken by the Solar Dynamics Observatory (SDO), Solar and Heliospheric Observatory (SoHO), and Solar-Terrestrial Relations Observatory (STEREO) to improve our understanding of how shock waves evolve in 3D during the eruption of a CME. We use our catalogue to search for relations between the shock wave's kinematic properties and the flaring activity associated with the underlying genesis of the CME piston.}
    {Past studies have shown that shock waves observed from multiple vantage points can be aptly reproduced geometrically by simple ellipsoids. The catalogue of reconstructed shock waves provides the time-dependent evolution of these ellipsoidal parameters. From these parameters, we deduced the lateral and radial expansion speeds of the shocks evolving over time. We compared these kinematic properties with those obtained from a single viewpoint by SoHO in order to evaluate projection effects. Finally, we examined the relationships between the shock wave and the associated flare when the latter was observed on the disc by considering the measurements of soft and hard X-rays.}
    {We find that at around 25 solar radii (R$_\odot$), the shape of a shock wave is very spherical, with a ratio between the lateral and radial dimensions (minor radii) remaining at around $b/a \approx 1.03$ and a radial to lateral speed ratio ($V_R/V_L$) $\approx 1.44$. The CME starts to slow down a few tens of minutes after the first acceleration and then propagates at a nearly constant speed. We revisit past studies that show a relation between the CME speed and the soft X-ray emission of the flare measured by the Geostationary Operational Environmental Satellite (GOES) and extend them to higher flare intensities and shock speeds. The time lag between the peak of the flare and of the CME speed is up to a few tens of minutes. We find that for several well-observed shock onsets, a clear correlation is visible between the derivative of the soft X-ray flux and the acceleration of the shock wave.}
    {}
    

\maketitle

\section{Introduction}
\label{sect_Introduction}

Coronal mass ejections (CMEs) are large-scale releases of plasma and magnetic fields expelled by the solar corona into interplanetary space \citep{Schwenn_2006, Rouillard_2011}, with a rate of occurrence directly linked to the level of solar activity \citep{Gopalswamy_2018}. A CME can drive a shock wave when its speed is greater than the characteristic speed of the ambient plasma, such as the fast magnetosonic speed \citep{Warmuth_2015}. Once the presence of CME is observed in coronal imagery, the properties of its shock wave can then be estimated \citep{Sheeley_2000, Ontiveros_2009, Bemporad_2010}. Shock waves are potentially efficient accelerators of solar energetic particles (SEPs) to high energy \citep[see for examples][]{Reames_1999, Kozarev_2015, Afanasiev_2018}, but the exact mechanisms involved are still under debate and stand as topics of active research \citep{Klein_Dalla_2017}. A better understanding of the structure and evolution of shock waves is therefore crucial to improving our understanding of the origin of SEPs, as well as space weather more generally \citep[see e.g. ][for recent studies]{Kouloumvakos_2020, Kouloumvakos2020b, Kouloumvakos_2022, Dresing_2022, Pesce-Rollins_2022}.\\

The advent of coronal and solar wind imaging from multiple vantage points with the Solar-TErrestrial Relations Observatory \citep[STEREO;][]{Kaiser_2008} opened up a new era in monitoring and modelling CMEs and their associated shock waves. In particular, this mission allowed for consistent multi-viewpoint imaging of the Sun and of the solar corona. In recent works, \citet{Rouillard_2016} and \citet{Kwon_2017} developed new techniques to track the 3D evolution of shock waves by fitting their geometry through the exploitation of different viewpoints. This type of 3D reconstruction provides the velocity of each point on the surface of the shock. When combined with numerical models of the background solar corona and solar wind, they can then be used to derive fundamental properties of the shock wave, such as the shock front speed, Mach number, or the angle of the shock normal with respect to the local magnetic field direction, $\theta_{BN}$ \citep{Rouillard_2016, Plotnikov_2017}. The measurement of $\theta_{BN}$ is important because it is suspected to have an impact on the efficiency of particle acceleration by a propagating shock front and could explain the variability in SEP compositions \citep{Tylka_2005, Tylka_2006}. \citet{Kouloumvakos_2019} improved and exploited this approach to model the evolution of 33 shock waves observed during the STEREO era that constitute the catalogue used in the present study. \\

An important aspect of fast and wide CMEs is their appearance in a so-called halo form. A halo CME is defined in \citet{Yashiro_2004} as a CME which appears to surround the occulting disc. Before STEREO, the occurrence of a halo CME was typically associated with an event propagating towards or at the 180 degree longitude line of the observing coronograph. Early studies based on imagery coming from multiple vantage points showed that powerful CMEs tend to produce strong coronal disturbances that appear, in some cases, to engulf the entire corona \citep{Rouillard_2011, Kwon_2015}. This global perturbation of the corona via the propagation of pressure waves can create the appearance of a halo-type signature at all observing platforms situated around the Sun.
The work of \citet{Kwon_2015} refined the definition of a halo CME by folding this fact. They show that 66\% of halo CMEs seen from Earth between 2010 and 2012 were also seen as halos by the STEREO spacecraft when they were situated at very different longitudes than the Earth. The key observation here is that the halo signature is not limited to the effect of material directly adjacent to the underlying magnetic flux rope, but includes also the outermost front associated pressure wave and shocks \citep{Kwon_2014}. The strong eruptive events exploited in the present study taken from the catalogue of \citet{Kouloumvakos_2019} are, in fact, all categorised as halo CMEs in the Solar and Heliospheric Observatory Large Angle and Spectrometric Coronagraph \citep[SoHO/LASCO;][]{Brueckner_1995} CME catalogue \citep{Yashiro_2004}.\\

\citet{DalLago_2003} exploited single vantage-point observations of halo CMEs by SoHO to infer the radial to lateral speed ratio of CMEs and found $V_{rad} = 0.88 \times V_{exp}$, with $V_{exp}$ the expansion speed of the CME. We will revisit the results of this analysis in light of our advanced catalogue of triangulated fast CMEs which removes the effect of the plane-of-sky projection \citep{Cremades_2004, Temmer_2009} inherent in studies based on single viewpoints. \\

We also revisit past studies that investigated the link between the kinematic evolution of CMEs and flaring activity. Coronal mass ejections are often associated with large flares (corresponding to X-class), especially the fastest, and conversely $>90\%$ of large flares are accompanied by CMEs \citep{Yashiro_2005}. Numerous studies have attempted to clarify the links between flares and CMEs \citep{Schmieder_2015}, by comparing CME kinematics with soft X-ray flux \citep{Maricic_2007, Salas-Matamoros_2015}, and hard X-ray flux \citep{Zhang_2001, Temmer_2010}. However, the exact physical relationship between these phenomena is complex and not fully understood \citep{Emslie_2012}.\\

\citet{Zhang_2004} found a clear relation between hard X-ray emission and the acceleration phase of CMEs, while \citet{Salas-Matamoros_2015} studied the link between CME kinematics and associated soft X-ray emission and found a correlation between these two with a Pearson correlation coefficient (CC) of 0.48. The soft X-rays are of thermal origin, produced by electrons heated during the flaring processes, whereas the hard X-rays are produced by non-thermal electrons and their effects are mostly observed during the impulsive phase of the soft X-ray emission \citep{Forbes_2000}. We revisit and extend these analyses by comparing the shock dynamics and the flare activity. This allows us to derive updated mathematical relations between these two classes of phenomena, shock kinematics, and flare, with potential space-weather applications. \\ 

This paper is structured as follows. In Sect. \ref{sect_Data_Methods}, we present the data and methods for analysing the 32 CMEs' shock wave and their kinematics. The analysis of shock shape and kinematics is presented in Sect. \ref{sect_ShockWave} which includes an analysis of the effects of projection. Section \ref{sect_shock_XR} revisits and extends past studies of the link between CME-shock kinematics and solar flares. Our conclusions for our results and their limitations are discussed in Sect. \ref{sect_discussion}.

\section{Data and methods}
\label{sect_Data_Methods}

\begin{figure}[h]
    \centering
    \includegraphics[scale=0.7]{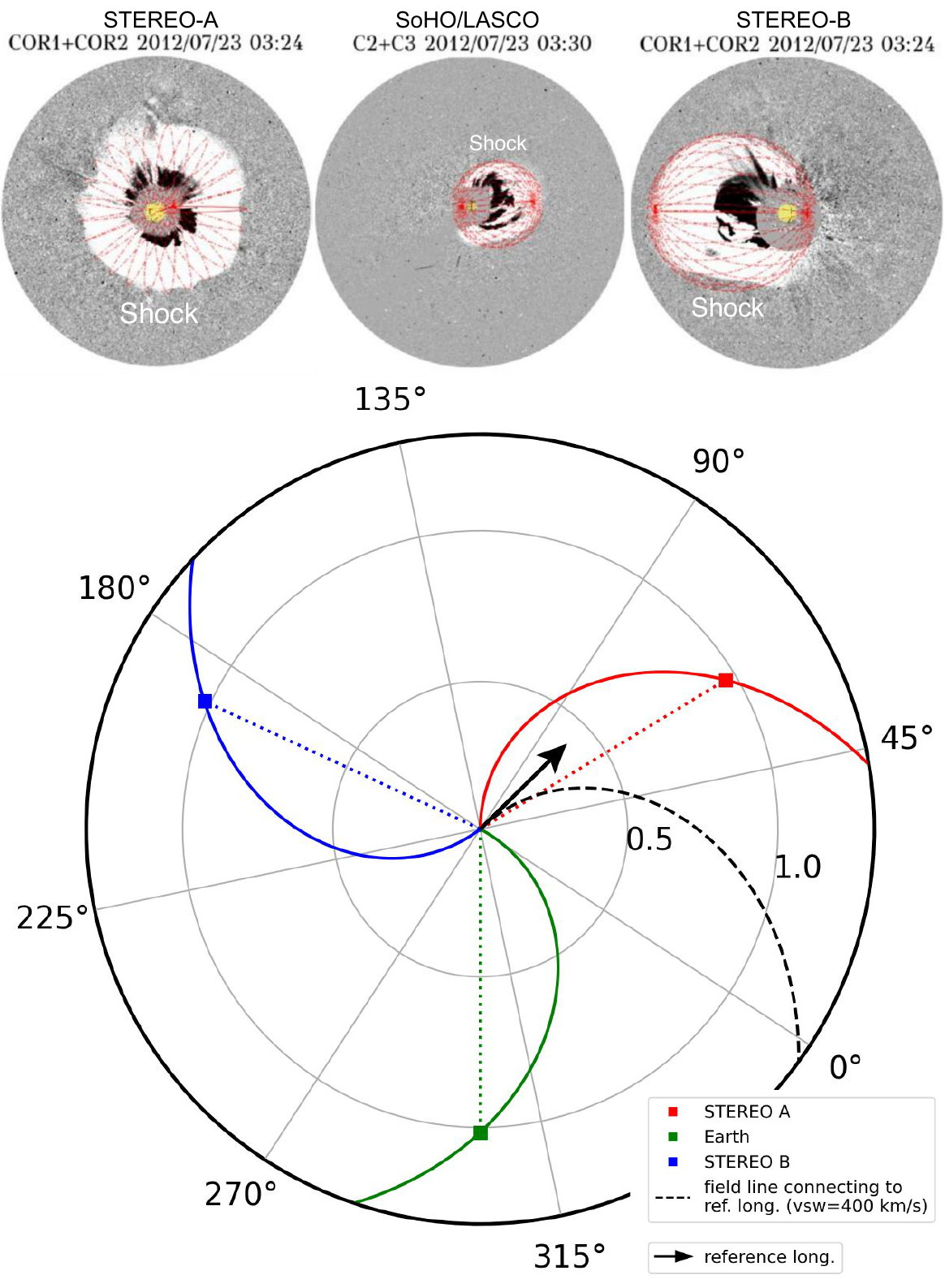}
        \caption{Representation to describes shock-wave triangulation using three different viewpoints, with the 23 July 2012 event. Top: Three points of view running difference images of a halo CME in white-light, with STEREO-A, SoHO/LASCO, and STEREO-B (shown from left to
right). Red dashed lines show the shape of a reconstructed shock wave. Bottom: View of the ecliptic from the solar north that shows the spacecraft position around the Sun at the CME eruption time. The direction of propagation of the CME is shown by the black arrow. This panel is produced using Solar-MACH online tool (available on \url{https://solar-mach.streamlitapp.com/}).}
    \label{3d_reconstruction}
\end{figure}

\citet{Kouloumvakos_2019} selected 33 CMEs between 2011 and 2017 that produced strong pressure waves in the solar corona during their eruption. Another important selection criterion was that these pressure waves had to be associated with significant concomitant SEP events. The events considered were typically observed simultaneously by several imagers on SoHO/LASCO and the two STEREOs (STEREO-A and STEREO-B). Thanks to these multiple vantage points, they applied their technique to reconstruct the time-evolving 3D ellipsoidal shape of each pressure wave to create a catalogue of 3D shock properties. Figure \ref{3d_reconstruction} presents an example of the shock-fitting technique applied to the powerful 23 July 2012 event by exploiting combined STEREO and SoHO data.\\

We now present an analysis of the statistical properties of 32 out of the 33 triangulated shock waves. One was removed from the study because the 3D fit carried some significant uncertainty reflected as strong discrepancies with the properties listed in the SoHO/LASCO CME catalogue \citep{Yashiro_2004, Gopalswamy_2009}.
Our sample of 32 events consists of rather extreme cases of fast and wide CMEs and is not representative of the average statistics of all CMEs in the SOHO/LASCO CMEs catalogue \citep{Yashiro_2004}, described in Table \ref{cme_stat}.

\renewcommand{\arraystretch}{1,2}
\begin{table}[h]
\centering
    \caption[]{Statistics on the speed and width of the CMEs listed in the SoHO/LASCO CME catalogue \citep{Yashiro_2004}, corresponding to CMEs recorded between the 11 January 1996 and the 31 August 2022. The standard deviation $\sigma$ is without units.}
    \begin{tabular}{ c || c | c | c || c }
          & max & median & mean & $\sigma$ \\ \hline \hline
         speed (km/s) & 3387 & 318 & 376.4 & 240.0 \\ \hline
         width (degree) & 360 & 35 & 54.1 & 63.9 \\ \hline 
    \end{tabular}
    \label{cme_stat}
\end{table}

Speed is measured in kilometres per second. The mean of CME linear speeds is around 380 km/s, while in our sample, the mean of CME speeds is around 1450 km/s, according to the same catalogue. This is because the selection of these events made by \citet{Kouloumvakos_2019} was such that they all included clear pressure or shock waves in the corona and large SEP events.
Associated CME shock speeds in the present pool of events range from 1070 km/s to 3600 km/s with a median of 2000 km/s. When considering the subset of CMEs associated with observed flares, we find that their mean class is X1.95, according to the Geostationary Operational
Environmental Satellite (GOES) measurements. The properties of these 32 triangulated shock waves deduced from our analysis are listed in Table \ref{tab_events}, by date and time of an event onset serving as an official reference. Columns 3 and 4 compare the CME speed from the SoHO/LASCO catalogue with the mean speed of the shock wave from the \citet{Kouloumvakos_2019} catalogue.

\begin{table*}[t]
\caption[]{List of studied events. The location of the eruption on the Sun surface is given in columns 5 and 6 in Stonyhurst heliographic coordinates, with the mean latitude $\langle HGLT \rangle$ and mean longitude $\langle HGLN \rangle$ in degrees. In this system of coordinates, the zero point is set at the intersection of the Sun's equator and its central meridian as seen from the Earth, which allows us to have a more accurate idea of the direction of a CME, compared to the direction of the Earth. Thus, we classified CMEs in three categories on column 7: on disc, far side, and limb, depending on the longitude of the CME. Between -90 and +90, the eruption occurs on the visible surface of the Sun, and therefore noted on disc. On the opposite side. The class of the associated soft X-ray flare measured by GOES is given for CME events originating on the disc, and the availability of the Reuven Ramaty High Energy Solar Spectroscopic Imager \citep[RHESSI;][]{Lin_2002} data for the hard X-ray flux is listed as 'complete', 'OK', or 'incomplete' depending on the quality of the data. Finally, the number of sunspots on the solar surface the day of the event is given to provide a reference of the solar activity.}
\footnotesize
\begin{tabular}{ c  c  |  c  c  |  c  c  c  |  c  c  | c}

\hline \hline ~Date~ & ~Time~ & CME Speed $ ^a$ & $\langle V_R \rangle$ $^b$ & $\langle HGLT \rangle$ $^b$ & $\langle HGLN \rangle$ $^b$ & ~location~ & GOES Class $^c$ & RHESSI data & day sunspots \\ \hline

15 Feb 2011 & 01:56 & 669.4 & 814.0* & -18.61 & 10.06 & on disc & X3.3 & complete & 100\\
7 Mar 2011 & 20:12 & 2125.4 & 2005.8 & 34.07 & 49.80 & on disc & M3.7 & complete & 122\\
21 Mar 2011 & 02:00 & 1341.1 & 1234.9 & 22.45 & 135.48 & far side & ... & ... & 34\\
4 Aug 2011 & 03:57 & 1315.1 & 1973.9 & 19.13 & 38.54 & on disc & X1.3 & ... & 81\\
6 Sep 2011 & 22:20 & 575.0 & 985.7* & 16.69 & 17.14 & on disc & X3.0 & incomplete & 93\\
22 Sep 2011 & 11:01 & 1904.6 & 1922.6 & -5.35 & -79.42 & on disc & X2.1 & ... & 86\\
4 Oct 2011 & 09:23 & 388.9 & 1051.2 & 36.90 & -157.16 & far side & ... & ... & 126\\
3 Nov 2011 & 22:20 & 991.1 & 956.1 & 4.93 & -154.84 & far side  & ... & ... & 149\\
23 Jan 2012 & 03:59 & 2174.7 & 2148.0 & 35.60 & 18.14 & on disc & X1.2 & ... & 108\\
27 Jan 2012 & 18:37 & 2507.8 & 2209.6 & 31.88 & 77.60 & on disc & X2.5 & ... & 39\\
5 Mar 2012 & 04:09 & 1530.5 & 1538.7 & 21.65 & -51.00 & on disc & X1.6 & ... & 105\\
7 Mar 2012 & 00:24 & 2684.4 & 2629.9 & 19.33 & -32.21 & on disc & X7.7 & ... & 102\\
24 Mar 2012 & 00:20 & 1152.0 & 1556.6 & 19.96 & -178.56 & far side & ... & ... & 65\\
17 May 2012 & 01:47 & 1581.8 & 1533.9 & -1.02 & 79.13 & on disc & M7.3 & ... & 114\\
23 Jul 2012 & 02:20 & 2003.2 & 2423.0 & 1.09 & 134.73 & far side & ... & ... & 60\\
20 Sep 2012 & 15:00 & 1201.7 & 2195.8* & -26.22 & -148.54 & far side & ... & ... & 68\\
27 Sep 2012 & 23:57 & 947.3 & 1067.3 & 7.65 & 33.70 & on disc & C5.4 & ... & 97\\
5 Mar 2013 & 03:00 & 1316.0 & 1268.7 & 2.48 & -142.45 & far side & ... & ... & 106\\
22 May 2013 & 13:32 & 1466.3 & 1331.9 & 16.39 & 76.64 & on disc & M7.2 & OK & 107\\
5 Oct 2013 & 07:00 & 963.6 & 995.5* & -28.85 & -116.9 & far side & ... & ... & 69\\
11 Oct 2013 & 07:25 & 1200.2 & 995.5 & 5.12 & -100.10 & far side & ... & ... & 115\\
25 Oct 2013 & 08:01 & 586.9 & 733.0* & -3.42 & -72.00 & on disc  & X2.5 & OK & 148\\
28 Oct 2013 & 15:15 & 811.5 & 1189.3 & 10.31 & -30.43 & on disc & M6.3 & OK & 155\\
2 Nov 2013 & 04:35 & 827.6 & 1114.1* & -5.48 & 152.02 & far side & ... & ... & 123\\
7 Nov 2013 & 10:30 & 1404.8 & 1858.2* & -5.24 & -143.24 & far side & ... & ... & 159\\
28 Dec 2013 & 17:30 & 1118.4 & 932.0 & 3.55 & 118.96 & far side & ... & ... & 95\\
6 Jan 2014 & 07:45 & 1401.9 & 1408.6 & 0.24 & 109.77 & far side & ... & ... & 245\\
7 Jan 2014 & 18:32 & 1830.4 & 2170.0 & -27.10 & 29.83 & on disc & X1.7 & ... & 196\\
25 Feb 2014 & 00:49 & 2146.5 & 1866.8 & -17.31 & -81.99 & on disc & X7.1 & OK & 157\\
1 Sep 2014 & 11:00 & 1900.5 & 1863.4 & -1.29 & -128.47 & far side & ... & ... & 94\\
10 Sep 2014 & 17:45 & 1267.4 & 1444.4 & 13.08 & 6.63 & on disc & X2.3 & ... & 161\\
10 Sep 2017 & 16:06 & 3162.9 & 2079.6 & -13.35 & 92.20 & limb & X11.9 & incomplete & 38\\ \hline
\end{tabular}
\footnotesize{\newline $^a$ From SOHO/LASCO catalogue, \\ $^b$ From \citet{Kouloumvakos_2019} catalogue \\ $^c$ New science-quality data available on https://satdat.ngdc.noaa.gov/sem/goes/data/science/xrs/, with a correction factor of 1.42 compared to operational data. \\ * Events which suffer from the lack of 3D tracking during the acceleration phase.}
\label{tab_events}
\end{table*}

Because of their high speed and of the acquisition rate of the STEREO instruments, some shock waves could not be fitted accurately before the nose of the structure reached a heliocentric distance of 2 R$_\odot$. This precludes inferring the CME acceleration phase, moreover, the estimation of the maximum speed of the shock wave was performed after this initial acceleration phase. Thus, 25 of the 32 events, for which the initial acceleration phase was captured accurately, were used later in the study, when we were considering the details of the CME kinematics soon after onset. The others are marked with a star in the $\langle V_R \rangle$ column.

\begin{figure}[t]
    \centering
    \includegraphics[scale=0.5]{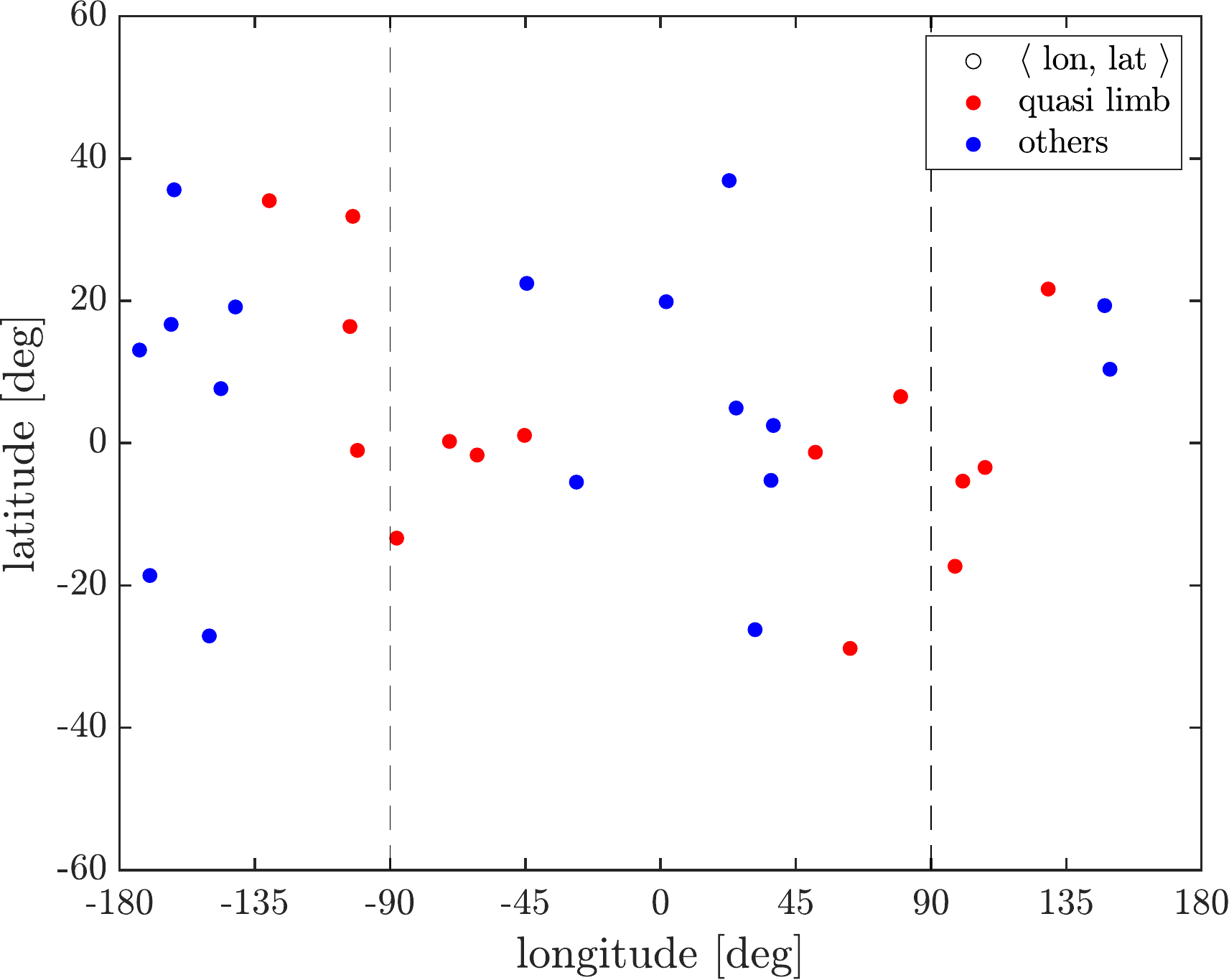}
        \caption{Figure representing the location of events projected on the Sun surface, with the latitude as a function of longitude in the Stonyhurst heliographic coordinate system. The circles locate the mean latitude and longitude for a given event, and are coloured in red for limb and quasi-limb event, i.e. with a longitude $\in [45^\circ,135^\circ] \cup [-45^\circ,-135^\circ]$, and in blue for the others. Black dashed lines represent delimitation between far side (two opposite side of the figure) and on disc (the middle of the figure) events.}
    \label{map_events}
\end{figure}

The heliocentric coordinates\footnote{The Stonyhurst heliographic system (HGS) is used, hence, the origin is the centre of the Sun, the z-axis is aligned with the Sun's north pole and the x-axis is aligned with the projection of the Sun-Earth line onto the Sun's equatorial plane, so (0$^\circ$, 0$^\circ$) is centred on the Earth's position.} of the central axis of these triangulated shocks are shown as filled circles in Fig. \ref{map_events}. Since the triangulated ellipsoid can shift slightly in latitude and longitude as the CME evolves in the corona, a circle represents the latitude and longitude of the central axis averaged over the duration of the event. The classification in two colours was chosen to separate the events into two distinct populations, allowing us thereafter to highlight the effects associated with the direction of propagation of CMEs relative to the observer (see Sect. \ref{sect_ShockWave_proj}). Red circles correspond to events propagating close to the plane of the sky as viewed from Earth, that is, with a longitude $\in [45^\circ, 135^\circ] \cup [-45^\circ, -135^\circ]$ in Stonyhurst heliographic coordinates. For simplicity, they are referred to as   'quasi-limb' in the rest of this article. In blue, the other events that propagate towards or away from the Earth with a longitude $\in [-45^\circ, 45^\circ] \cup [-135^\circ, -180^\circ] \cup [135^\circ, 180^\circ]$ in Stonyhurst heliographic coordinates.

The dashed lines represent the limb between on disc (at longitudes $-90^\circ$ and $90^\circ$) and far-side events. We see that the latitudinal distribution of events is limited to the active region belts inside a latitude band of $[-40^\circ, +40^\circ]$. This is expected since the events considered here are powerful CMEs that produced strong shock waves and SEPs, these solar storms tend to form in the direct vicinity of active regions. The catalogue provides a uniform distribution of events in longitude, allowing us to evaluate such effects as projections in coronal images.

\section{Geometrical properties of shock waves}
\label{sect_ShockWave}

\begin{figure}[h!]
    \centering
    \includegraphics[scale=0.5]{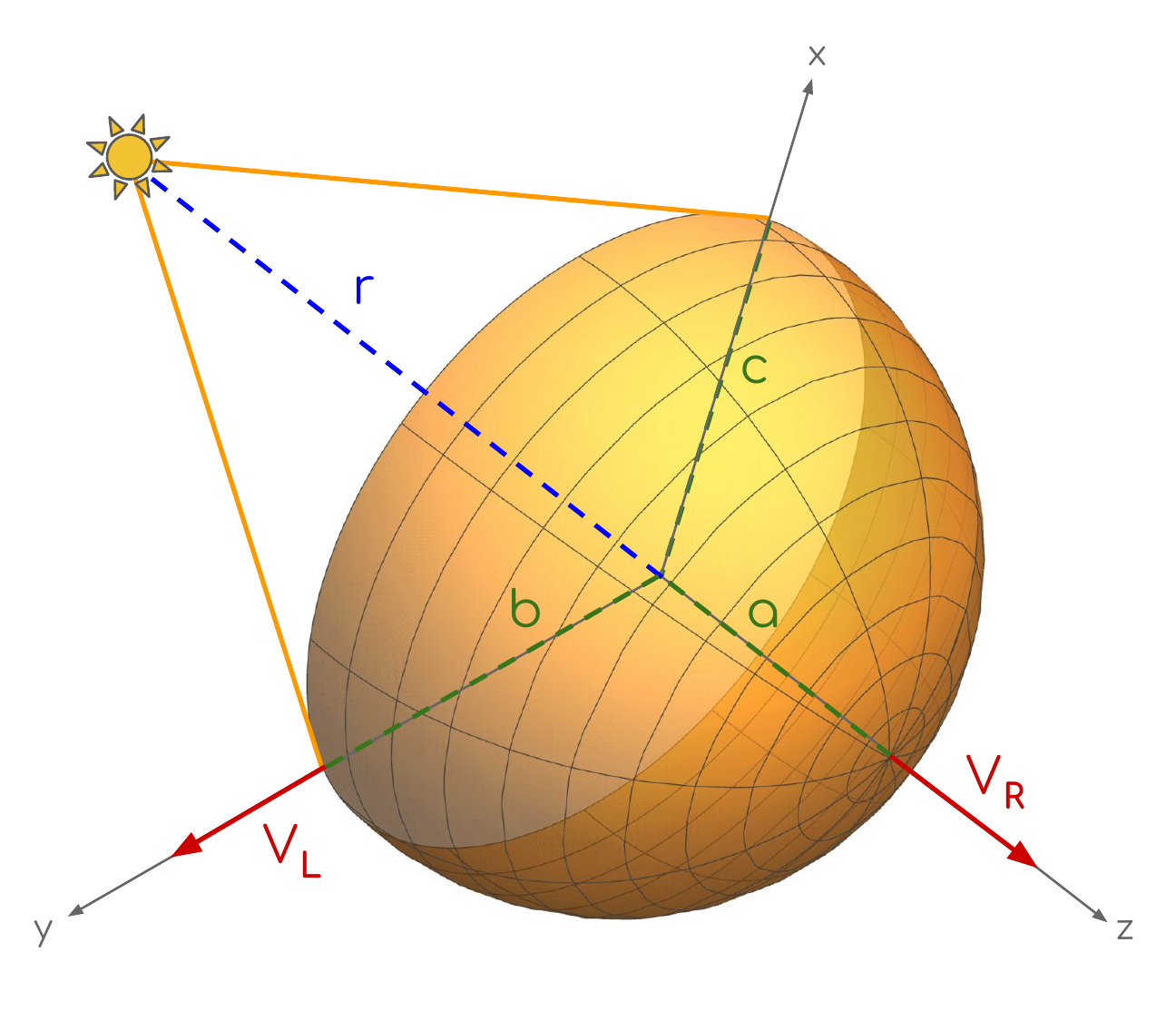}
        \caption{Schematics of the shock wave fit. The parameters displayed are those of the ellipsoid model: $r$ (blue dashed line) represents the distance between the surface of the Sun and the centre of the ellipsoid; $a$, $b,$ and $c$ (green dashed lines) are the half-axes; $V_R$ and $V_L$ (red line) represent  its radial and lateral speeds, respectively. The shock wave is represented by a half-sphere for illustrative purposes, i.e. $a(t) \approx b(t) \approx c(t) $ between 2 and 25 solar radii, which confirms the cone model.}
    \label{ellipsoid_schema}
\end{figure}

The fitted geometrical model, namely, the ellipsoid schematised in Fig. \ref{ellipsoid_schema}, has three half-major axes: $a(t)$ in the radial direction, and $b(t)$ and $c(t)$ for its cross-radial (or lateral) dimensions. By construction, north-south and east-west asymmetric expansion cannot be distinguished. The statistical study therefore does not reflect specific cases, only the overall structure of the shock. However, there are events with significant expansion asymmetries, as shown, for example, in \citet{Majumdar_2021}. We note, that our fitting methodology and statistical approach does not allow for this class of events to be highlighted.
By taking the time derivative of $R_S(t) = r(t)+a(t)$, the distance between the Sun and the nose of the CME-driven shock wave (which is called 'apex'), we can derive the radial speed $V_R(t)$, corresponding to the propagation speed of the shock wave in the solar corona. The lateral speed $V_L(t)$ corresponds to the expansion speed and is obtained by considering the lateral half major axis, $b(t)$, since the $c(t)$ follows closely the $b(t)$ variation in our modelling results. In the sample of 32 events, 7 suffer from the lack of 3D tracking during the first ten minutes of the eruption (identified with a star in the $\langle V_R \rangle$ column), the acceleration phase of the CME was not tracked with sufficiently accuracy in the low corona.

\subsection{Expansion ratio of 3D shock waves}
\label{sect_ShockWave_geo}

\begin{figure}[h!]
    \centering
    \includegraphics[scale=0.35]{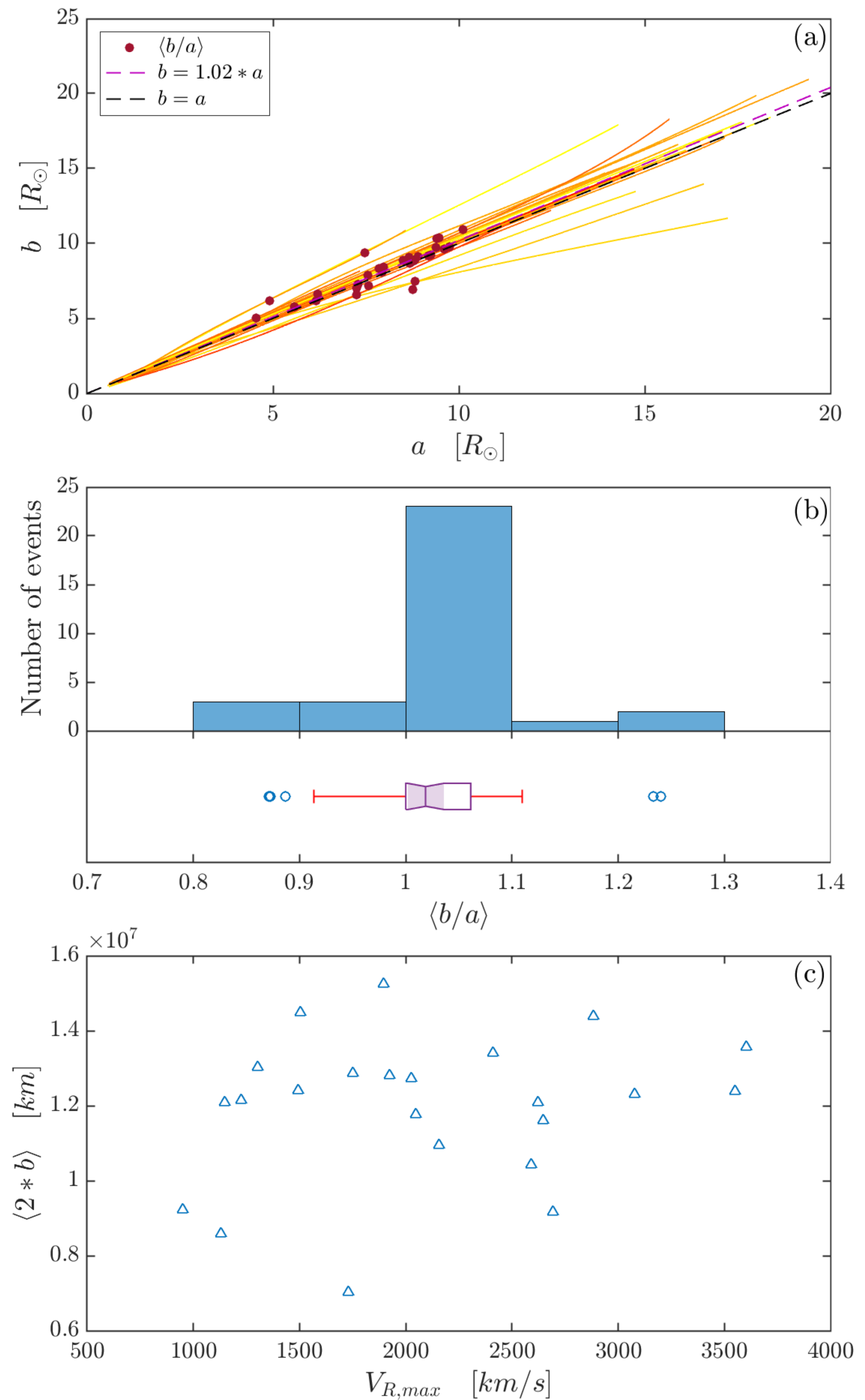}
        \caption{Figure representing the geometry statistics of the 32 shock waves.
        Panel (a): Ellipsoid parameter $b$ as a function of ellipsoid parameter $a$ for the 32 events. The colours correspond to the solar activity of the event day, with, from yellow to red, the smallest to the largest number of sunspots observed. The red points are the average $b/a$ ratio for each event. The purple dashed line corresponds to the overall $\langle b/a \rangle$ ratio, i.e. $b = 1.02 \times a$ and the black dashed line to a case where $b = a$.
        Panel (b): In two parts. Top : Histogram of $\langle b/a \rangle$ ratio. Bottom: Boxplot of $\langle b/a \rangle$ ratio. On the purple box, the central mark indicates the median, whereas the bottom and top edges indicate the $25$th and $75$th percentiles, respectively. The whiskers extend in red to the most extreme data points, and in dot blue are plotted individual data points considered outliers.
        Panel (c): Mean width of the shock wave, $\langle 2  b \rangle$, as a function of $V_{R,max}$, the maximum radial speed of the shock wave.}
    \label{fig_b_vs_a}
\end{figure}

The catalogue gives the time evolution of the ratio $b(t)/a(t)$, which represents the ratio between the longitudinal width of the shock wave and its radial extent. This provides direct information on how the shape of shock waves evolves dynamically over time. Panel (a) of the Fig. \ref{fig_b_vs_a} represents the half axis $b(t)$ as a function of the half axis $a(t)$ in units of solar radii (R$_\odot$), with points corresponding to the average values for each event. The distance between the shock wave nose and the surface of the Sun, $R_S(t)$, evolves between 2 and 25 R$_\odot$. This upper limit corresponds to the distance beyond which shock triangulation becomes impossible with white-light coronagraphs alone. Each of the 32 events corresponds to a line coloured from yellow to red related to the number of sunspots measured on the day of the CME event with high sunspot numbers corresponding to the red lines and small sunspot numbers shown as yellow lines. There is no clear relation between the solar activity level and the evolution of the $b(t)/a(t)$ ratio. Past studies have suggested that the extent of CMEs observed in coronagraphs could be related to the general level of solar activity with weaker cycles marked by smaller CME widths \citep{Gopalswamy_2014}. We could not test this idea here since our sample only covers one solar cycle period, and is strongly biased by high speeds and large widths; however, we do note that under these conditions, there is no clear dependence on the activity level.

\begin{figure}[h!]
    \centering
    \includegraphics[scale=0.3]{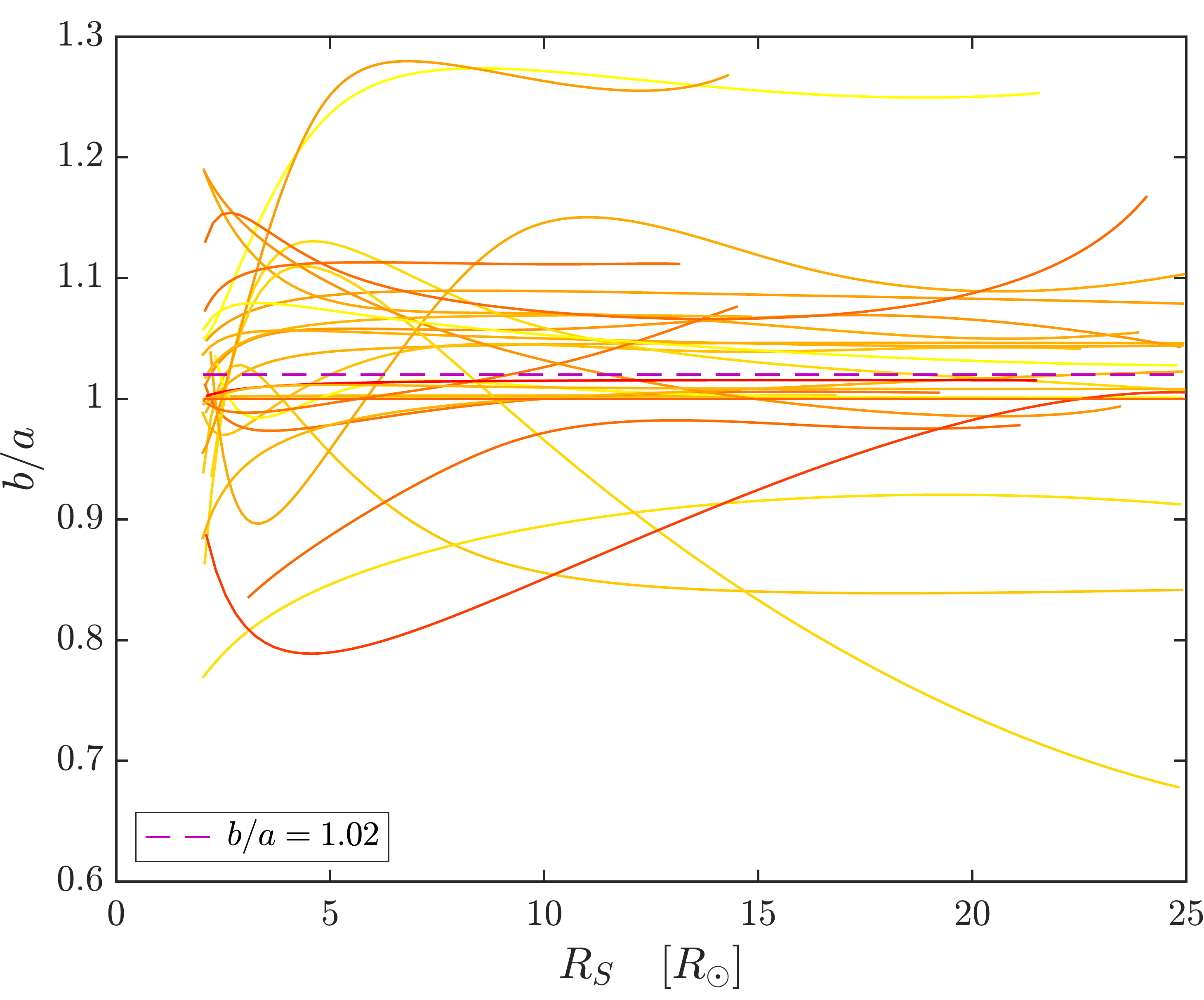}
        \caption{Ratio $b(t)/a(t)$ as a function of $R_S$ in solar radius units for the 32 events. The colours correspond to the solar activity of the event day, with, from yellow to red, the smallest to the largest number of sunspots observed. The purple dashed line represents the overall $\langle b/a \rangle$ ratio, i.e. $b = 1.02 \times a$.}
    \label{fig_ba_vs_r}
\end{figure}

For each event, the mean ratio $\langle b/a \rangle$ is noted by a red dot, while the global ratio average including all events is $\langle b/a \rangle = 1.02$, represented by a purple dashed line. Three events present a constant $b(t)/a(t) = 1$ during the propagation represented by black dashed lines. Figure \ref{fig_ba_vs_r} shows the evolution of the ratio $b(t)/a(t)$ as a function of R$_\odot$ for the 32 events; 21 out of the 32 events have a $b(t)/a(t)$ that increases during propagation, whereas 8 events see their $b(t)/a(t)$ decrease during propagation; and 27 events out of 32 have a ratio that stabilizes to a constant value around 20 R$_\odot$, and 22 of them with a value between 1 and 1.1.

Panel (b) of Fig. \ref{fig_b_vs_a} combines a histogram (top) and a boxplot (bottom) of the $\langle b/a \rangle$, the mean ratio $b(t)/a(t)$ for the 32 modelled shock ellipsoids. The average is taken in the range of radial distances $R_S(t)$ between 2 and 25 R$_\odot$. The histogram peaks at $\langle b/a \rangle$ between 1 and 1.1 for 23 of 32 events ($72\%$), as also observed in the boxplot. For $81\%$ of the shock waves $\langle b/a \rangle$ is between 0.92 and 1.12, and half of them have $\langle b/a \rangle$ between 1 and 1.05. The average of the overall sample is $ \langle b/a \rangle = 1.03 \pm 0.08 $, with the error corresponding to the standard deviation.

The panel (c) of the same figure shows an estimation of the shock wave lateral width $\langle 2  b \rangle$ as a function of the maximum radial speed $V_{R,max}$. As explained above, via a modelling hypothesis, we cannot distinguish lateral asymmetries different from those affecting $b$ with respect to $c$, the two semi-major axes of the shock. In this case, $\langle 2  b \rangle$ remains a correct estimation of the shock width. The comparison between the shock wave lateral width, $\langle 2  b \rangle,$ and the maximum radial speed, $V_{R,max}$, does not reveal a clear dependence between these two. We do not find a relation between the widths of CMEs as seen in coronographic images and their heliocentric radial speed.

\subsection{Shock waves kinematics}
\label{sect_ShockWave_kin}

\begin{figure}[h!]
    \includegraphics[scale=0.35]{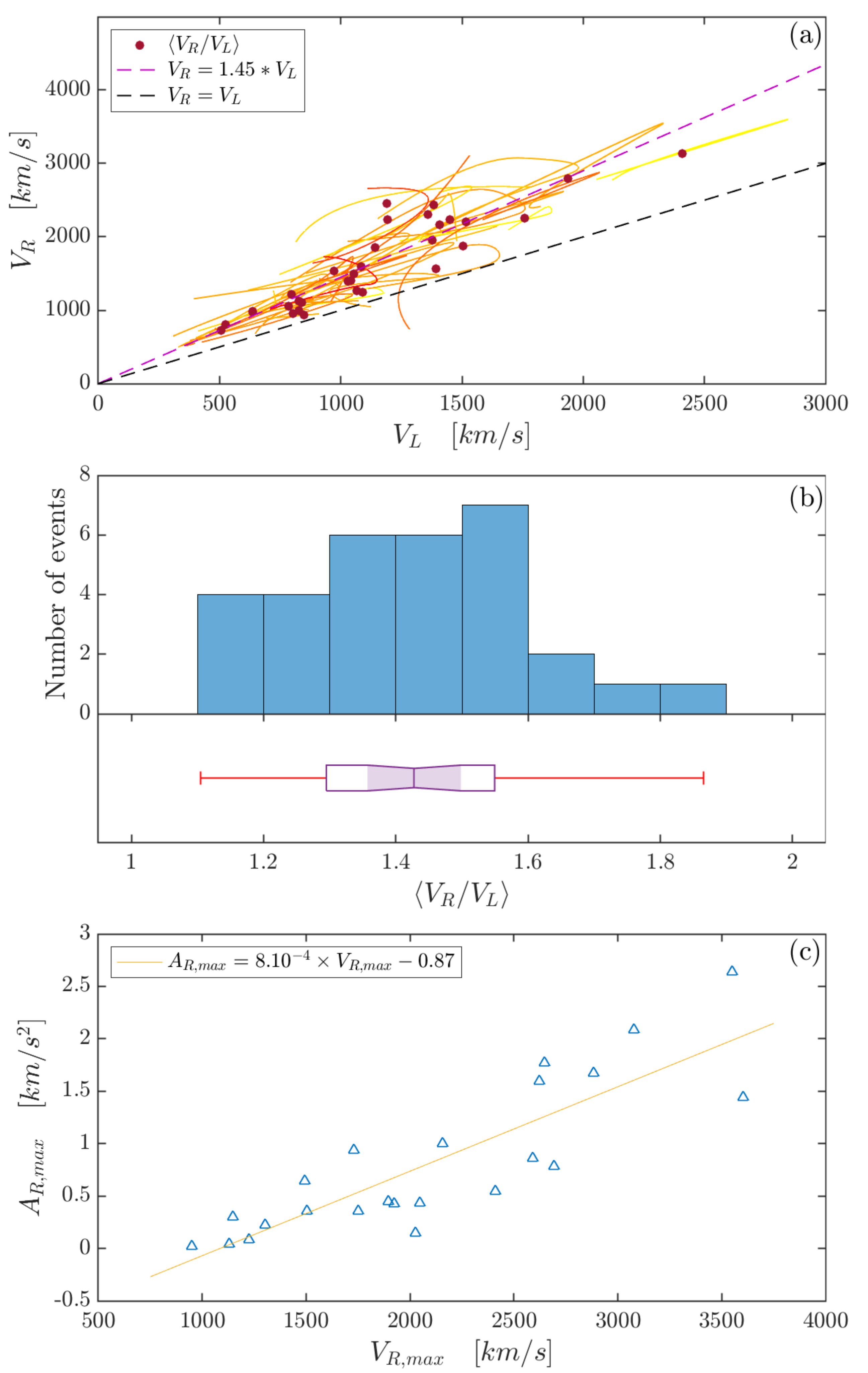}
        \caption{Figure representing the kinematic statistics of the 32 shock
        waves.
        Panels (a) and (b) are the same as panels (a) and (b) in Fig. \ref{fig_b_vs_a}, but instead for radial speed, $V_R$, and lateral speed, $V_L$.
        Panel (c): $A_{R,max}$ as a function of $V_{R,max}$ for 25 shock waves (on 32) presenting a visible acceleration phase. Each point of colour represents a different event. The yellow line corresponds to the best fit obtained from these data.}
    \label{fig_vr_vs_vl}
\end{figure}

Panel (a) of Fig. \ref{fig_vr_vs_vl} presents the radial speed ,$V_R(t),$ as a function of the lateral speed $V_L(t)$. The dashed black line marks the limit of $V_R(t) = V_L(t)$ and we see that the shock radial speed $V_R(t)$ is always greater than the lateral speed, $V_L(t)$. Nevertheless, the rather spherical evolution of the shock shown in the previous section implies that $V_L(t)$ remains elevated throughout the propagation of the shock wave to 25 $R_\odot$. Panel (b) of the same figure represents a summary of $\langle V_R/V_L \rangle$ ratio observations. The values are predominantly located between 1.3 and 1.6, with the following relation:

\begin{equation}
    \langle V_R/V_L \rangle = 1.44 \pm 0.22\,.
    \label{eq_vr_vl}
\end{equation}

This relation can be compared with those obtained for CMEs \citep{DalLago_2003, Gopalswamy_2009b, Shen_2013}, as discussed in Sect. \ref{sect_discussion}.

The ellipsoid centre is moving away from the Sun at the same time that the structure itself is expanding as a sphere, with $a \approx b$, as shown in Sect. \ref{sect_ShockWave_geo}. The outward speed of the ellipsoid centre (see Fig. \ref{ellipsoid_schema}) is $dr/dt = V_R - da/dt$, thus on the order of 0.3 - 0.6 $V_L$ (according to Eq. \ref{eq_vr_vl}).\\

Panel (c) of Fig. \ref{fig_vr_vs_vl} shows the relation between the maximum radial acceleration and the maximum radial speed (blue triangles), for the 25 out of 32 CMEs shock waves presenting an acceleration phase (see Sect. \ref{sect_Data_Methods}). The yellow line corresponds to the best fit for these data:

\begin{equation}
    A_{R,max} = (8 \pm 2).10^{-4} \times V_{R,max} - (0.87 \pm 0.50)\,,
    \label{eq_armax_vrmax}
\end{equation}

with a CC of 0.84. The speed $V_R$ was obtained by using two-point derivative (leap-frog method) of the position $X_R$ . The acceleration $A_R$ was obtained by deriving $V_R$ in the same way.

\subsection{Effect of projection on derived CME kinematics}
\label{sect_ShockWave_proj}

\begin{figure}[t]
    \centering
    \includegraphics[scale=0.46]{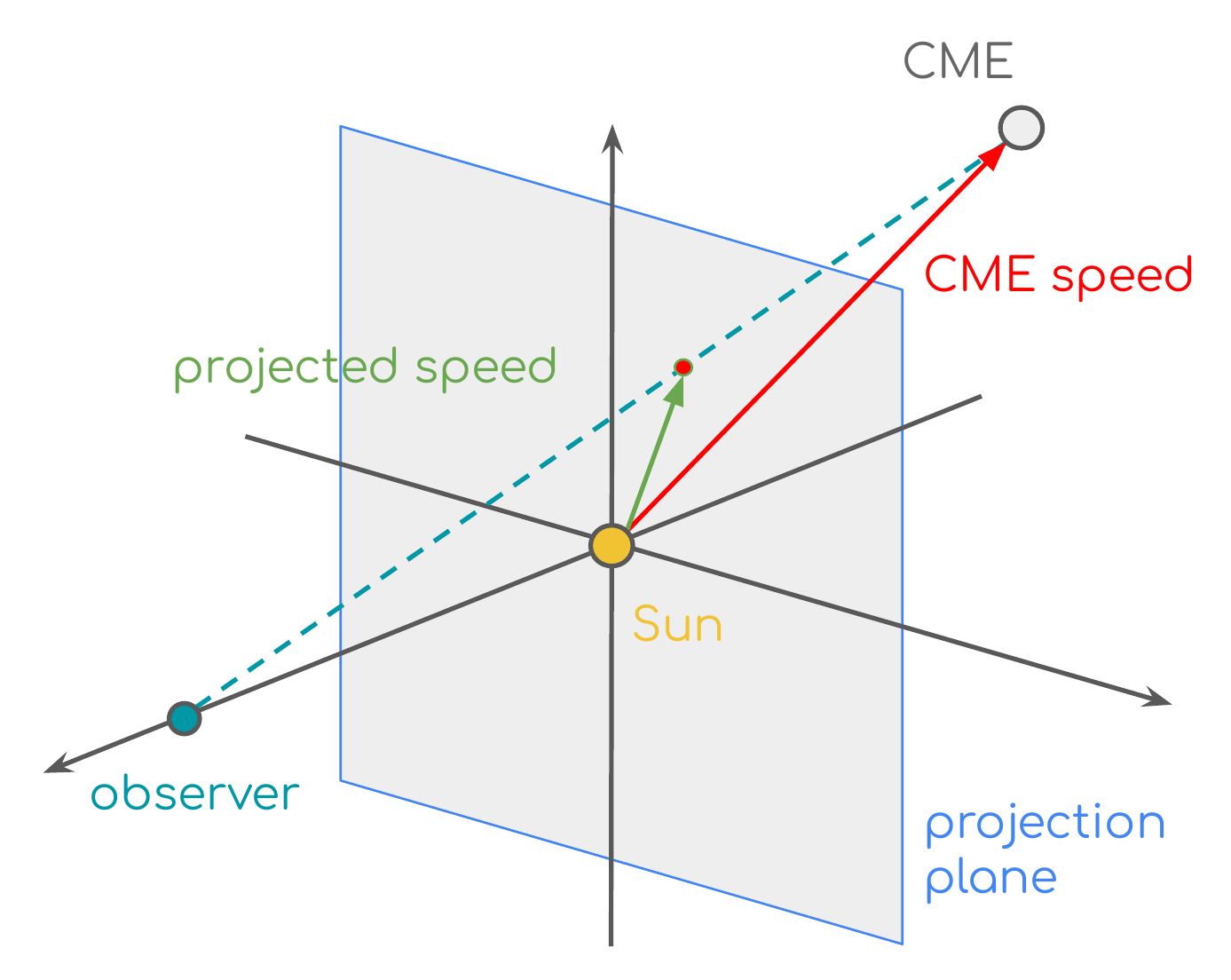}
        \caption{Schematic view of a CME event. The red arrow represents the real velocity vector while the green arrow is the projection on the plane-of-sky plane as seen from an observer from Earth (i.e. L1 or SoHO/LASCO) point of view.}
    \label{fig_projection_effect}
\end{figure}

One of the most complete CME catalogues based on white-light images is the SoHO/LASCO catalogue mentioned previously \citep{Yashiro_2004}. This catalogue provides CME speeds calculated only from one vantage point, that of the SoHO telescope at L1 point. This leads to well-known limitations associated with plane-of-sky projection effects \citep{Burkepile_2004} schematised in Fig. \ref{fig_projection_effect}. To limit any projection effects, many studies reduce their sample sizes by considering only limb CMEs \citep{Schwenn_2005}, that is, CMEs propagating in the plane of the sky seen of the observing instrument. From these events, CME kinematics were derived in coronagraph images close to the Sun and not far out with say heliospheric imagers \citep{Rouillard_2011}. For CME events associated with well-defined flux ropes, more involved forward-modelling approaches, such as the graduated cylindrical shell (GCS) model, can also provide de-projected speeds regardless of the propagation direction of a CME \citep{Shen_2013, Thernisien_2009}. The GCS model provides a shock spheroid model structure to also track the shock front.\\

\begin{figure}[h!]
    \centering
    \includegraphics[scale=0.3]{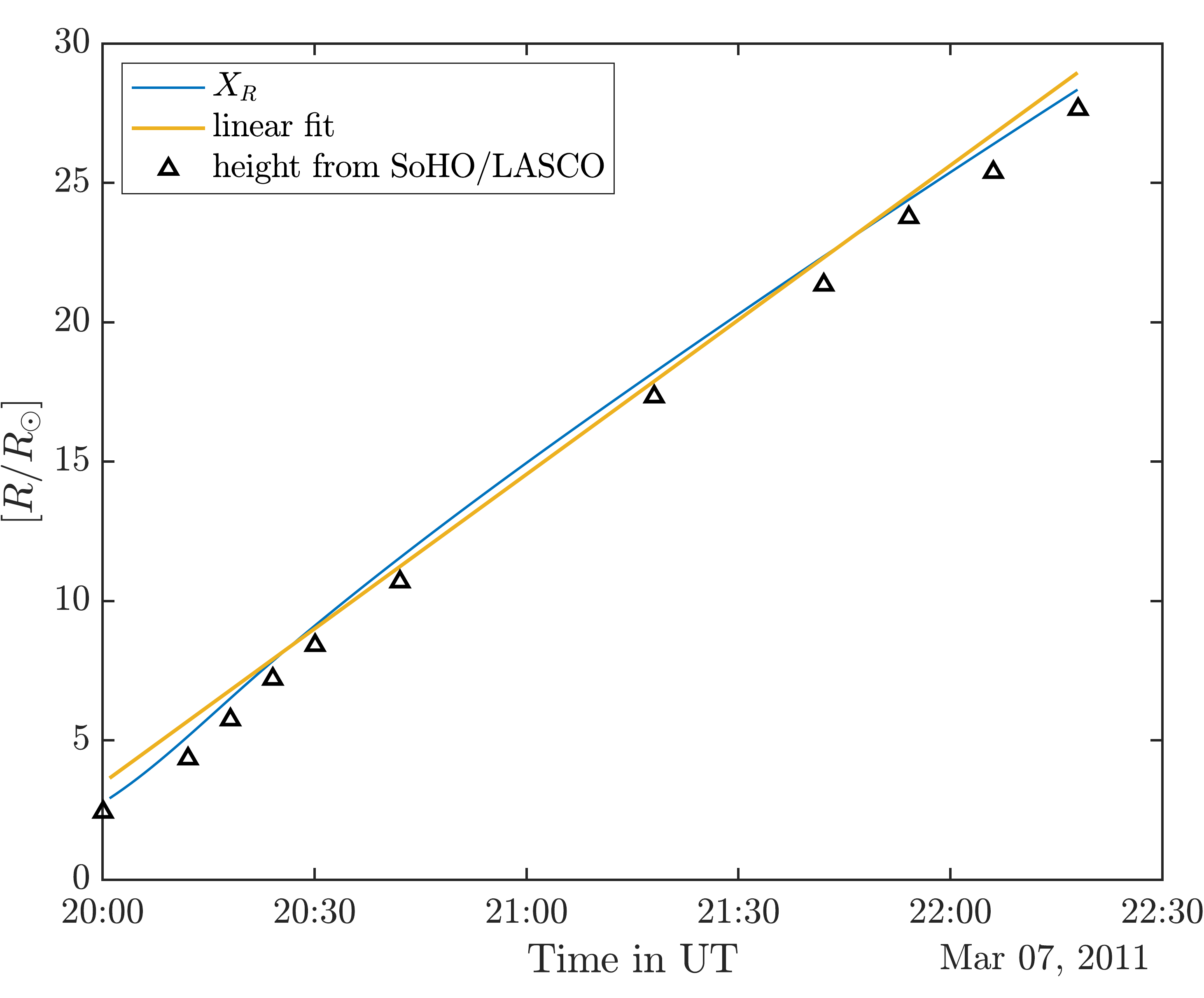}
        \caption{Figure representing data for the 7 March 2011 CME event. Shock wave apex position, $X_R,$ (in blue) and linear fit of this position (in yellow) as a function of time. Black triangles are from the SoHO/LASCO CME catalogue for the same event \citep{Yashiro_2004}.}
    \label{fig_position_20110307}
\end{figure}

We can estimate the effects of projection by comparing the SoHO/LASCO CME catalogue with the kinematics derived from our catalogue of 3D shock waves. We have chosen to base ourselves on this catalogue because it is the most used to date for space weather related topics. Figure \ref{fig_position_20110307} presents the height as a function of time for the 7 March 2011 event. Black triangles are measurement points of CME height from SoHO/LASCO catalogue. Blue line corresponds to the 3D shock wave apex position ($X_R$), and has been cut according to the start and end of the measurements of SoHO/LASCO. The yellow line is the linear fit of the 3D position $X_R$. This fit gives a constant speed $V_{R,linear}=1773.6$km/s, while the SoHO/LASCO fit, resulting from tracing a single point on the CME leading edge, gives $V_{LASCO} = 2146.5$km/s.

\begin{figure}[h!]
    \centering
    \includegraphics[width=\hsize]{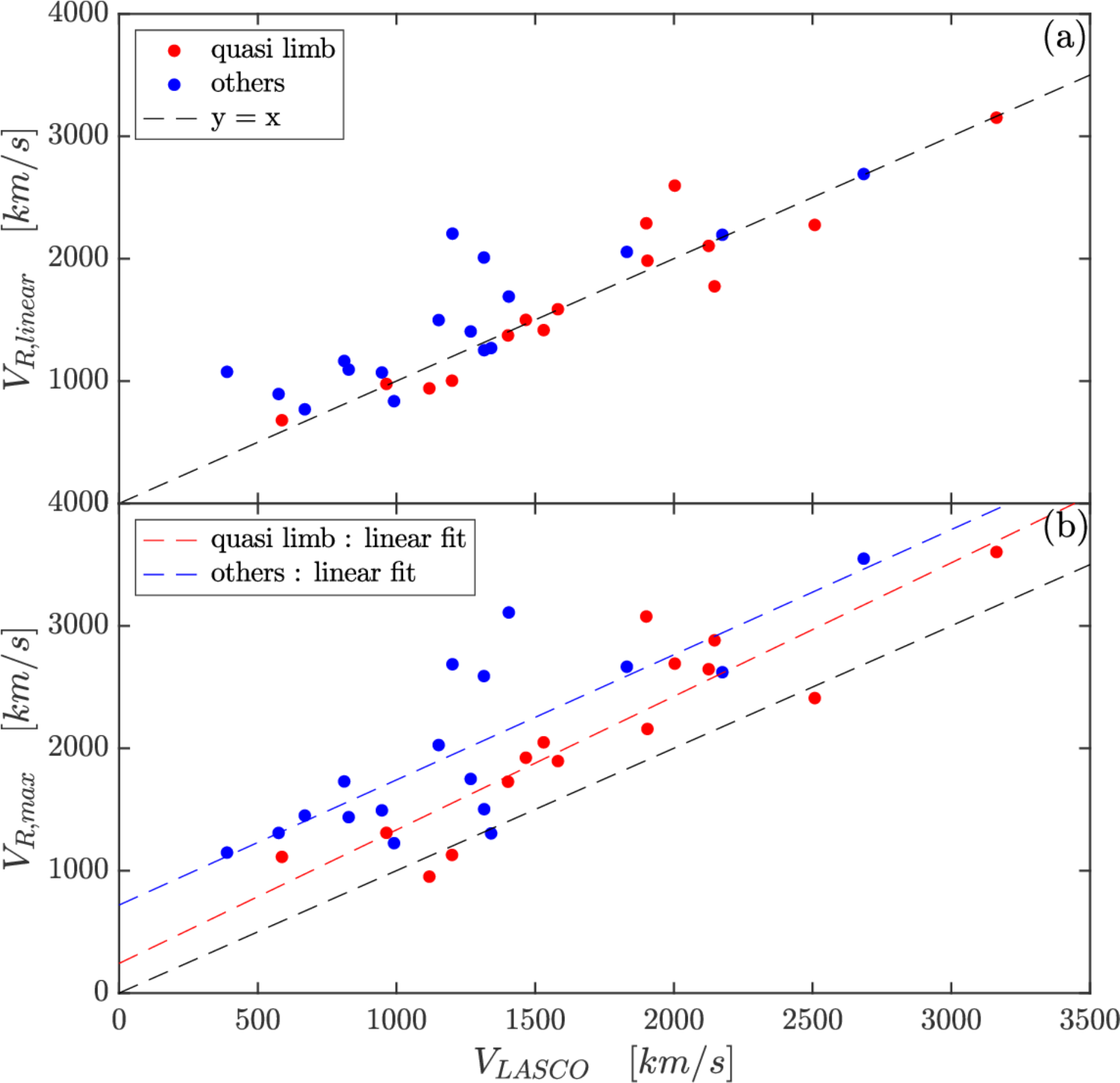}
        \caption{Figure representing the projection effects as a function of $V_{LASCO}$, which is the linear CME speed from SoHO/LASCO CME catalogue.
        Panel (a): $V_{R,linear}$, the radial speed calculated from the derivation of the linear fit of position $X_R$ (yellow line in Fig. \ref{fig_position_20110307}). Red points correspond to events close to the limb, which are events at a longitude of $\in [45°,135°] \cup [-45°,-135°]$ in the Stonyhurst coordinate system. The blue points correspond to the others, with a longitude $\in [-45°,45°] \cup [-135°,-180°] \cup [135°,180°]$, directed  towards or inwards the Earth, respectively.
        Panel (b): $V_{R,max}$ with the same colour code. The red dashed line is a linear fit of the red points, corresponding to Eq. \ref{eq_limb_vlasco} with a CC of 0.89, and the blue dashed line is a linear fit of the blue points, corresponding to Eq. \ref{eq_nolimb_vlasco} with a CC of 0.78.}
    \label{fig_speeds_and_projections}
\end{figure}

A more systematic comparison between $V_{LASCO}$ and $V_{R,linear}$ is now possible by performing the same analysis with all 32 events. We divided our catalogue into two categories related to the propagation direction of the shock apex relative to the observer (see Sect. \ref{sect_Data_Methods} and Fig. \ref{map_events} for more details).

Panel (a) of Fig. \ref{fig_speeds_and_projections} represents $V_{R,linear}$ as a function of $V_{LASCO}$ for the 32 events with the previously described colours. Visually, red dots are closer to the black dashed line, representing the case where $V_{R,linear} = V_{LASCO}$ than the blue dots. In general, there is a tendency to underestimate $V_{LASCO}$ compared to $V_{R,linear}$. This confirms the existence of a projection effect, less important for quasi-limb events. The spread of the data is too large to give a general correction factor for this effect, and it strongly depends on the position of the event on the Sun's surface seen from the Earth at eruption time. Panel (b) of the same figure represents $V_{R,max}$, the original 3D maximal speed from \citet{Kouloumvakos_2019}, as a function of $V_{LASCO}$ with the same colour code. There is an important difference between the two sets of speeds with, at another time, a clear underestimation of $V_{LASCO}$. Only four events (21 Mar 2011, 27 Jan 2012, 11 Oct 2013, and 28 Dec 2013) present a $V_{LASCO}$ higher than $V_{R,max}$, with a difference of less than 170 km/s. The latter may be due to measurement uncertainties on $V_{LASCO}$ or on the 3D modelling itself. The two dashed lines in red and blue are respectively linear fit of quasi-limb events (red points) and no-limb events (blue points). They give us Eqs. \ref{eq_limb_vlasco} and \ref{eq_nolimb_vlasco} with CCs of 0.89 and 0.78.

\begin{equation}
    V_{R,max} = (1.09 \pm 0.36) \times V_{LASCO,~ql} + (242.00 \pm 572.20)
    \label{eq_limb_vlasco}
,\end{equation}
\begin{equation}
    V_{R,max} = (1.02 \pm 0.43) \times V_{LASCO,~nl} + (719.90 \pm 576.10)\,.
    \label{eq_nolimb_vlasco}
\end{equation}

In these equations, $ql$ is used to define quasi-limb and limb events, whereas $nl$ is used for the other non-limb events.
To conclude this part of the study, the apex speed of the CME seen by SoHO/LASCO ($V_{LASCO}$) is underestimated compared to the maximum of the 3D radial speed of the shock wave ($V_R$). However, a correction factor applied according to their location on disc could be taken into account to mitigate this effect.

\section{Association between shock properties and X-ray flares}
\label{sect_shock_XR}

We go on to focus on the subset of shock waves that were associated with flares observed on the visible disc viewed from Earth and study the relation between these shock waves and the flares. One goal was to search for a possible relation between soft and hard X-ray (SXR and HXR) emissions and the kinematic properties of the induced CME-driven shock.

\begin{figure}[h!]
    \centering
    \includegraphics[width=\hsize]{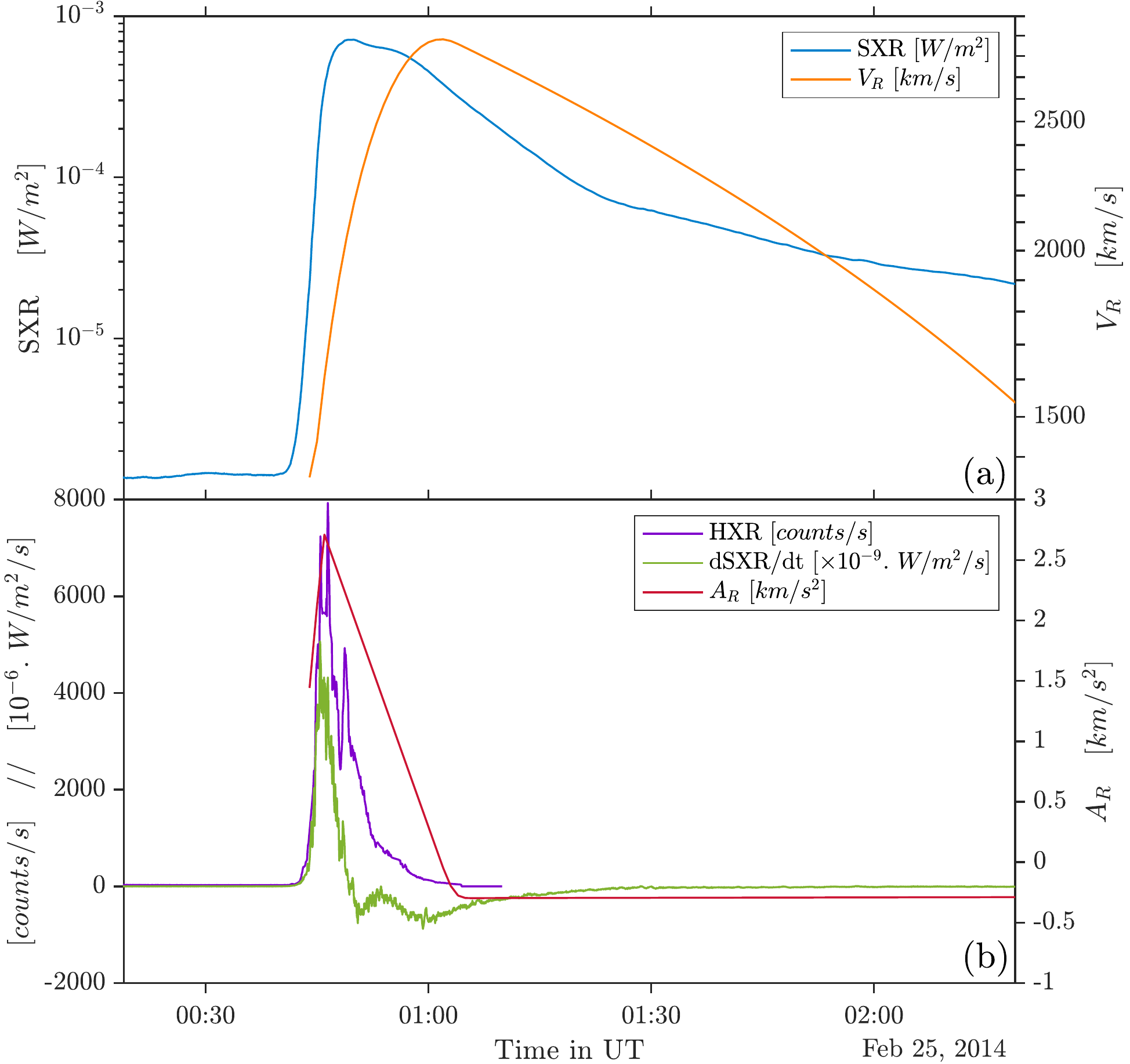}
        \caption{Different X-ray flux and kinematic parameters as a function of time for 25 Feb 2014 event.
        Panel (a) represents, on a logarithmic scale, the soft X-ray (SXR) flux in blue in $10^{-7}.W/m^2$ on the left, and the radial speed $V_R$ in orange with units in $km/s$ on the right.
        Panel (b) represents, on a linear scale, to the left, the hard X-ray (HXR) flux in purple in $counts/s$ and the soft X-ray derivative (dSXR/dt) flux in green in $10^{-6} W/m^2/s$. On the right: Radial shock wave acceleration, $A_R$, in $km/s^2$ with the red line.}
    \label{fig_records_20140225}
\end{figure}

Past studies \citep{Zhang_2001, Zhang_2004, Temmer_2010, Salas-Matamoros_2015} have found clear relations between the kinematic parameters of CMEs, such as their speed or acceleration, and flare measurements in SXR and HXR. In what follows, we revisit these past studies by comparing our de-projected CME kinematics with direct soft and hard X-ray measurements. \\

\subsection{Relations between soft- and hard- X-ray flares}

There is an important distinction between the SXR fluxes measured by GOES and the HXR count-rates recorded by the Reuven Ramaty High Energy Solar Spectroscopic Imager \citep[RHESSI;][]{Lin_2002}. GOES reflects the thermal emission of the hot flare plasma\ and, consequently, the peak GOES flux is strongly correlated with the maximum thermal energy \citep[see for example][]{Warmuth_2016}. In contrast, the HXR flux at higher energies (say above 25~keV) is dominated by non-thermal thick-target emission \citep{Brown_1971} and thus reflects the instantaneous flux and energy input by accelerated electrons. In many flares, we observe the Neupert effect \citep{Neupert_1968}, where the HXR emission corresponds to the derivative of the SXR emission. The physical reason for this is that the SXR-emitting thermal plasma is generated by the energy input due to the energetic electrons, which produce HXRs. The GOES derivative is thus often used as a proxy for the non-thermal HXR emission.

Among our shock wave sample, all on-disc events and one limb event (10 Sep 2017) were associated with X-ray flares measured by GOES. This corresponds to 18 shock waves listed with their associated flare classes in Table \ref{tab_events}. Considering that  $>90\%$ of large flares are accompanied by CMEs \citep{Yashiro_2005} and that all the fast CME events in our catalogue triggered strong shocks, the absence of a flare detection is almost certainly related to the flare-CME release taking place on the far side of the Sun. This is confirmed by the triangulation work since the estimated source regions of all CME events not associated with observed flares in Table \ref{tab_events} originate on the far side of the Sun as viewed from Earth. \\

Among the 18 events for which a soft X-ray flux was observed and associated with a CME, there were 8 that also occurred with hard X-ray flares detected by RHESSI and associated with the impulsive phase of the flare. The energy bands we consider from RHESSI are 25-50 keV and 50-100 keV, which should both be dominated by non-thermal emissions, except for very large X-class flares where 25-50 keV can also be contaminated by a super-hot thermal component. In these cases, we considered only the higher energy band, 50-100 keV, while for the other slightly weaker flares (7 Mar 2011, 22 May 2011,  Oct 2013), we considered the combined band 25-100 keV.\\

As an illustrative example, Fig. \ref{fig_records_20140225} presents a comparison of the soft and hard X-ray fluxes for the 25 Feb 2014 CME event. The panel (a) compares the hard X-ray (HXR, left axis) in purple with the soft X-ray (SXR, left axis) in blue. The units are respectively the counts/s for the HXR and $10^{-7}$.W/m$^2$ for the SXR, in logarithmic scale. The panel (b) compares the same HXR with the derivative of the soft X-ray flux (dSXR/dt, left axis) in green. The peak of the SXR flux and the peak of the HXR flux occur at similar times, with, for our height events, the SXR peak always occurring after the HXR peak. The delay between these two peaks is usually between 90 seconds and 20 minutes, with seven out of eight values under 10 minutes. For six out of the eight events, the peak in dSXR/dt flux occurred before the HXR peak, and for all of them the time difference is less than 12 minutes. Relations between HXR and SXR are as already documented in \citet{Veronig_2005}.

\begin{figure}[h!]
    \centering
    \includegraphics[width=\hsize]{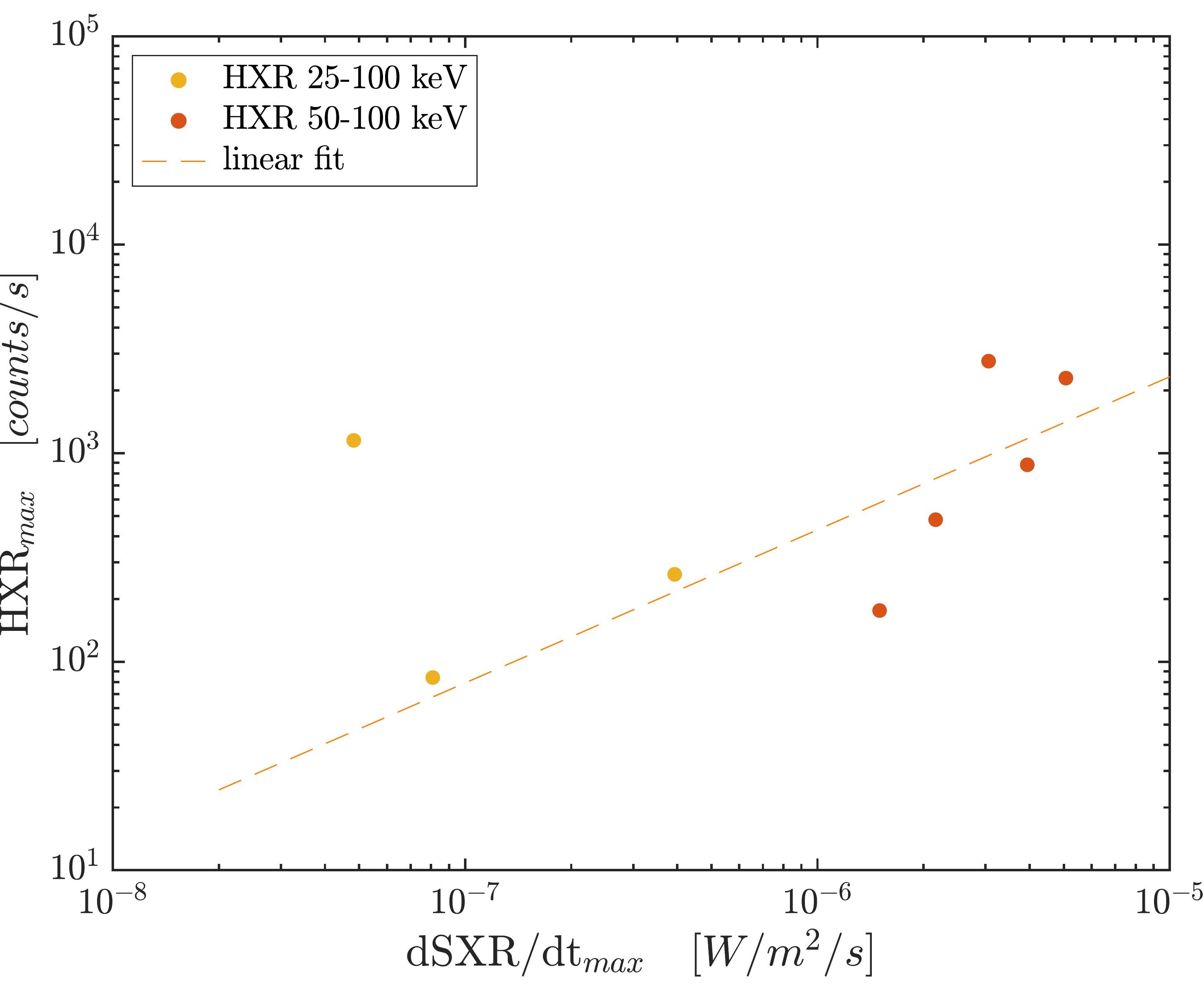}
        \caption{Peak of the HXR emission as a function of the peak of the soft X-ray derivative (dSXR/dt) flux for the height CMEs events with an associated hard X-ray flare. Yellow points are associated with HXR in the energy band 25-100 keV and red points to HXR in the energy band 50-100 keV because they are X-class GOES SXR. The orange dashed line represents the best fit for the data where we excluded the extreme yellow point with high HXR$_{max}$ and low dSXR/dt$_{max}$. The corresponding Eq. \ref{eq_hxr_dsxr} have a CC of 0.80.}
    \label{fig_hxr_vs_dsxr}
\end{figure}

Figure \ref{fig_hxr_vs_dsxr} represents for each event the peak of the hard X-ray flux (HXR$_{max}$) as a function of the peak of the soft X-ray derivative flux (dSXR/dt$_{max}$), for the height events associated with RHESSI detections of hard X-rays. Red and yellow circles represent the maximum of the HXR flux in the 25-100 keV and 50-100 keV bands, respectively. The best fit between HXR$_{max}$ and (dSXR/dt$_{max}$) is shown as a dashed line with:

\begin{equation}
    \log_{10}(\mathrm{HXR}_{max}) = (0.73 \pm 0.56) \log_{10}\bigg(\frac{\mathrm{dSXR}}{\mathrm{d}t}_{max}\bigg) + (7.04 \pm 3.29)\,.
    \label{eq_hxr_dsxr}
\end{equation}

The CC between HXR$_{max}$ and $\frac{\mathrm{dSXR}}{\mathrm{d}t}_{max}$ is $rr=0.80$. The event of 7 Mar 2011 was discarded of the fit because of the very small amplitude of its soft X-ray flare and the lack of GOES two second data on this period which prevents an accurate estimation of its maximum SXR flux.

This good correlation between the peak of the HXR emission and the one of the SXR (SXR$_{max}$) could be therefore be exploited as a substitute in case of bad or no-detection of the SXR flux. As an example, the event rejected from the fit because of its uncertain SXR flux has a measured dSXR/d$t_{max}$ of 4.82$\times 10^{-8}$. From the HXR$_{max}$ recorded, we calculate an expected dSXR/d$t_{max}$ of 3.55$\times 10^{-6}$.
We can too notice the presence of a trend for the 18 events associated with a flare. An impulsive flare, that is, with a strong rising slope, reveals a high SXR maximum, while a slight slope associated with a low dSXR/d$t$ peak is related to a low SXR peak.

\subsection{Relations between flare and shock wave kinematics}

Figure \ref{fig_records_20140225} also compares the temporal evolution of the shock wave and flare during the 25 Feb 2014 CME event. We show in panel (a): the HXR, SXR and the radial shock speed, $V_R$; and in panel (b): HXR, $\frac{\mathrm{dSXR}}{\mathrm{d}t}$, and the radial acceleration $A_R$.

As already mentioned in Sect. \ref{sect_Data_Methods}, the acceleration phase of 3 out of the 18 events could not be determined from the 3D shock fitting technique and was therefore not considered here. For the 15 other events, we observe that the maximum of radial speed, $V_R$, is always reached after the maximum SXR and that the two peaks have a temporal shift comprise between 5 and 32 minutes, such as in our example on panel (a) of Fig. \ref{fig_records_20140225}. The synchronisation between SXR and shock wave kinematics is easier to see in panel (b), which presents the time derivatives of the two aforementioned parameters, that is, the radial shock wave speed profile (in km/s, on the right) and $\frac{\mathrm{dSXR}}{\mathrm{d}t}$ (on the left). For events that could be tracked accurately during the early phase of the CME eruption, the peak of the radial acceleration is temporally correlated with the peak of $\frac{\mathrm{dSXR}}{\mathrm{d}t}$. Indeed, the two peaks occur at similar times but without precise order or offset, with a time difference between -5 and 10 minutes for all events; this result is  very close to that of \citet{Bein_2012}, except the ambiguous one of 28 Oct 2013 where two flares follow each other.
Panel (b) shows also close synchronisation between $A_R$ and the flare energy release in the hard X-ray (on the left), such as observations of \citet{Temmer_2008}. Among the eight events for which we recorded a HXR flux, there are the three mentioned above for which we could not track the acceleration phase due to the lack of needed observational data. Of the remaining five, three are strongly synchronized with the HXR peak with a time difference of less than 4 minutes, one occurs 15 minutes earlier and the last one 11 minutes later.

\begin{figure}[t]
    \centering
    \includegraphics[scale=0.3]{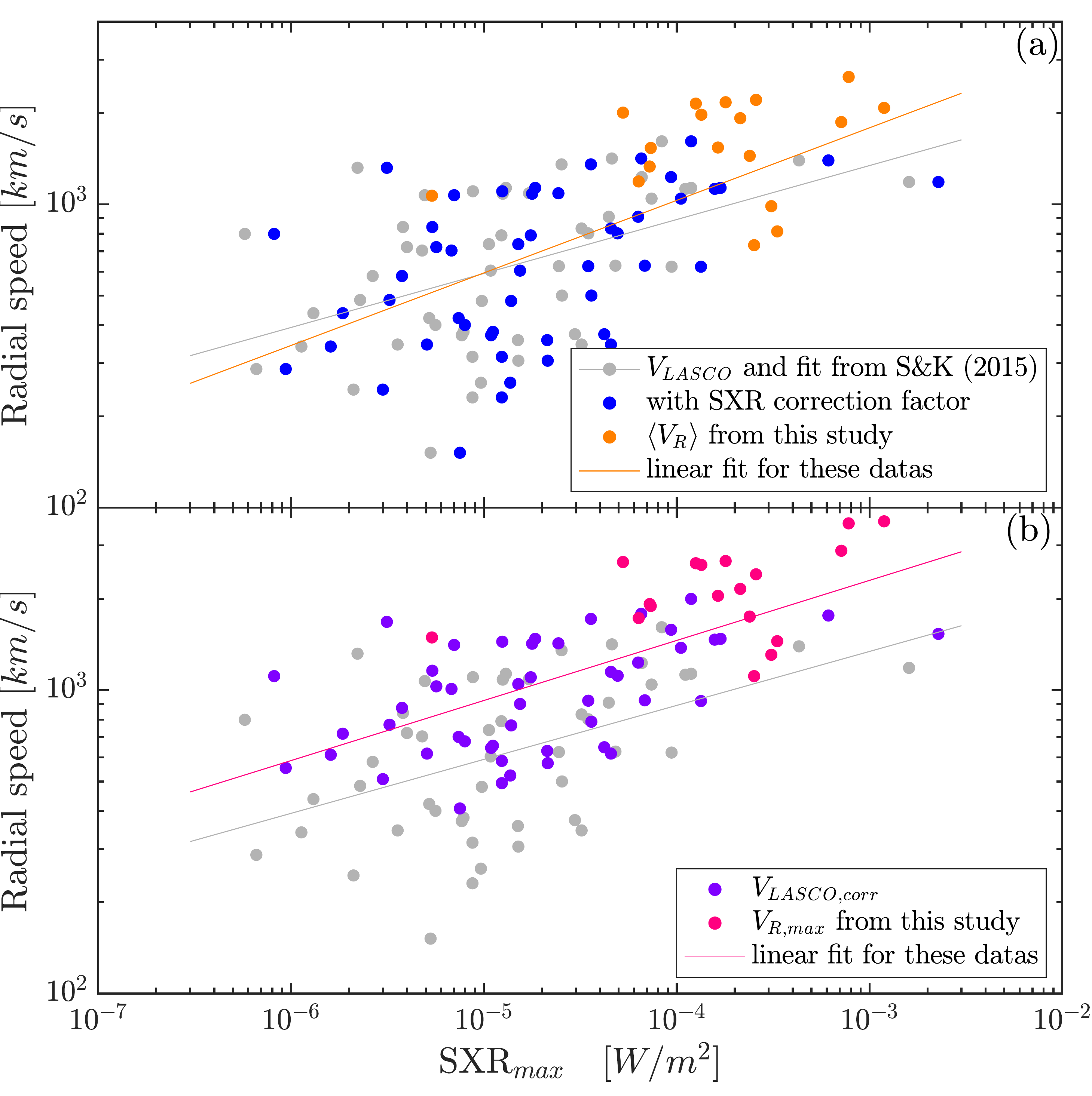}
        \caption{Figure representing radial speeds as a function of the SXR flux maximum (SXR$_{max}$). In the two panels, elements in grey come from Fig. 2 of \citet{Salas-Matamoros_2015}: dots represent CME speeds from SoHO/LASCO catalogue and the grey line is the best fit for them, given by Eq. \ref{eq_sm}. Panel (a): $\langle V_R \rangle$ (orange dots) from our sample and $V_{LASCO}$ (blue points) from \citet{Salas-Matamoros_2015} shifted because of the correction factor on SXR flux, according to GOES new science-quality data. The orange line corresponds to Eq. \ref{eq_vr_mean_sxr} and is the best fit for blue and orange points with a CC of 0.62.
        Panel (b): With pink dots, $V_{R,max}$ from our sample and with purple dots, $V_{LASCO, corr}$. $V_{LASCO, corr}$ is the same as precedent $V_{LASCO}$, but with a correction on the speed of quasi-limb events according to Eq. \ref{eq_limb_vlasco} to recover an estimation of the real 3D radial speed of the CME. The pink line corresponds to Eq. \ref{eq_vr_max_sxr} and is the best fit for purple and pink points with a CC of 0.65.}
    \label{fig_vr_vs_sxr}
\end{figure}

A statistical study presented by \citet{Salas-Matamoros_2015} considered 49 CME events observed on disc and far from the limb (at longitudes relative to a central meridian less than 80$^\circ$) to avoid occultation-related issues. They found a clear relationship between the maximum of the soft X-ray flare and the speed of the CME from the SoHO/LASCO catalogue. Their results are shown in grey in Fig. \ref{fig_vr_vs_sxr}. Then, $V_{LASCO}$ measured as a function of SXR flux maximum is shown by the grey dots. The correlation they obtained (grey line) is given by Eq. \ref{eq_sm} and have CC of 0.48:

\begin{equation}
    \log_{10}(V_{CME}) = (0.20 \pm 0.08) \log_{10}(\mathrm{SXR}_{max}) + (3.83 \pm 0.38)\,.
    \label{eq_sm}
\end{equation}

Our analysis permits to revisit this study linking the radial velocity of the shock wave with SXR$_{max}$. We effectively increased the sample size and improved on their statistics, especially since our shock waves are associated with faster CMEs that are not well represented in the \citet{Salas-Matamoros_2015} study. Our sample is shown in orange in panel (a) and pink  dots in panel (b) in Fig. \ref{fig_vr_vs_sxr}.

Blue dots in panel (a) are the data sample of \citet{Salas-Matamoros_2015} but with the SXR maxima corrected by a factor of 1.42. This follows recommendations made by the new release of GOES science-quality data\footnote{\url{https://satdat.ngdc.noaa.gov/sem/goes/data/science/xrs/}, see Sect. \ref{sect_Data_Methods} for more information.}, whereas $\langle V_R \rangle$ from our data show using orange dots. The combination of $\langle V_R \rangle$ (orange dots) and shifted $V_{LASCO}$ (blue dots) is designated by $Vm$. The fit of $Vm$ is given by Eq. \ref{eq_vr_mean_sxr} (orange line) has a CC of 0.62:

\begin{equation}
    \log_{10}(Vm) = (0.24 \pm 0.07) \log_{10}(\mathrm{SXR}_{max}) + (3.96 \pm 0.33)\,.
    \label{eq_vr_mean_sxr}
\end{equation}

Considering that all selected CMEs of \citet{Salas-Matamoros_2015} are limb CMEs, we applied our new correction given by Eq. \ref{eq_limb_vlasco} to the 2D speed of SoHO/LASCO in order to estimate their real 3D radial shock speed. These corrected values are called $V_{LASCO,corr}$ and are represented by purple dots in panel (b). Equation \ref{eq_vr_max_sxr} is a fit with our $V_{R,max}$ data in pink, where $V$ represents the combination of $V_{LASCO, corr}$ and $V_{R,max}$:

\begin{equation}
    \log_{10}(V) = (0.20 \pm 0.06) \log_{10}(\mathrm{SXR}_{max}) + (3.91 \pm 0.25)\,.
    \label{eq_vr_max_sxr}
\end{equation}The CC of this relation is now 0.65.

\section{Summary and discussion}
\label{sect_discussion}

We exploited a catalogue of 32 triangulated shock waves to provide a statistical analysis of their kinematic and geometric evolution. The triangulation technique assumes that shock waves have an ellipsoidal geometry, this geometric assumption has provided a good description of the topological evolution of shock waves during the first hours of a CME's expansion as shown in many previous studies \citep{Kwon_2015, Kwon_2017, Rouillard_2016}. The main results of the present study are the following:

\begin{itemize}
\item  The ratio of a shock's transverse to radial extent is on average equal to $b/a = 1.03 \pm 0.08 $ in a range of heliocentric distances comprised between 2 and 25 R$_\odot$. This means that the first few hours of a shock wave's evolution are marked by a spherical expansion.

\item During its propagation to 25 R$_\odot$, the radial speed of a CME shock is linked the lateral speeds of the CME through the following relation: $V_R = (1.44 \pm 0.22) V_L$. This relation could prove useful for inferring the probable global 3D expansion of CME associated shock, even when it is observed from a single vantage point.

\item We found that there is a linear relationship between the maximum of the radial speed, $V_R,$ and the maximum of the radial acceleration, $A_R$ (see panel (c) of Fig. \ref{fig_vr_vs_vl}).

\item Projection effects as expected impact more the non-limb CMEs than the quasi-limb. Correction factors can be applied to derive the real 3D speed from the SoHO/LASCO speed (Eqs. \ref{eq_limb_vlasco} and \ref{eq_nolimb_vlasco}).

\item We confirm the well-known relation between the derivative of the soft X-ray flux and the hard X-ray flux during a flare known on the Neupert effect. We find that the temporal shift between the two peaks is less than 12 minutes. The maximum of the derivative of the soft X-ray flux and the maximum of the hard X-ray flux are related by Eq. \ref{eq_hxr_dsxr} with a CC of 0.80.

\item The correlation between the radial velocity of the shock wave and the maximum SXR flare improves with increasing maximum flare intensity. While there is a scatter of one order of magnitude in speed for flares of class M or less, this scatter becomes smaller for flares of class X. Furthermore, all X-class flares are associated with CMEs reaching speeds of at least 1000 km/s.

\end{itemize}

However, it is important to note the limitations of this study. Further out in the heliosphere, shock waves develop more complex shapes due to their interaction with the formed solar wind \citep{Wood_2012} and in the interplanetary medium, as seen in the results of \citet{Janvier_2013, Janvier_2015}. Due to their interactions with different solar winds, shock waves can slow down more rapidly in certain directions and develop pancake or even more complex shapes \citep{Wood_2012}. 

Furthermore, similarities and differences between our conclusions, detailed in \ref{sect_ShockWave_kin}, and the literature can be highlighted. \citet{DalLago_2003} for example used 57 limb CMEs and found a relation between the CME radial speed, $V_{rad}$, and the CME expansion speed $V_{exp}$ : $V_{rad} = 0.88 * V_{exp}$, similarly to the Eq. \ref{eq_vr_vl} derived in the present study. \citet{Kwon_2017} showed that in the case of halo CMEs, $V_{rad}$ is a good approximation for the CME shock wave speed $V_R$. However, $V_L$ is not a good approximation of $V_{exp}$ because of the physical differences between them. Thus, a direct comparison of the relationships between them is not possible.
However, \citet{Gopalswamy_2009b} found that the relation between the radial and expansion speeds of CMEs depends on the CME width. Considering an expansion speed of the shock, $V_{exp,shock} = 2*V_L$ for our sample, a decreasing trend in the ratio $V_R/V_{exp,shock}$ as the shock width $\langle 2  b \rangle$ increases is indeed observed, but we did not find a clear relation between the two.

Moreover, in Sect. \ref{sect_ShockWave_kin}, we obtained Eq. \ref{eq_armax_vrmax} between $A_{R,max}$ and $V_{R,max}$, assuming a linear relationship. In a recent study, \citet{Majumdar_2021b} derived kinematic properties of CMEs using the GCS approach. In particular, they also derived the relationship between $A_{max}$ and $V_{max}$ (of the CMEs). The relation linking all their sample is $A_{max} = 10^{-3.35}V_{max}^{1.21}$ km/s$^2$. We compared their $A_{max}$ to $A_{R,max}$ determined for the shocks and find that  $A_{max}$ is typically larger than $A_{R,max}$ by a factor of at least 3-5. We attribute this difference to different methodologies followed in these studies. A future study could focus on a selection of well-observed events and compare directly the kinematics of CME flux ropes derived from the GCS technique with the kinematic derived from the present shock fitting technique.

The relations derived in the present study are useful to address events for which the acceleration phase is missed due to limited data coverage. For example, we can recover an estimation of $A_{R,max}$ from $V_{R,max}$ thanks to the relation given in panel (c) of Fig. \ref{fig_vr_vs_vl}. If the maximum speed is missed, another possibility is to use $V_{LASCO}$ then Eqs. \ref{eq_limb_vlasco} and \ref{eq_nolimb_vlasco} from Fig. \ref{fig_speeds_and_projections}. An approximation of the real maximum 3D radial speed can be derived depending on the CME location on the Sun, and this one could be used to obtain $A_{R,max}$.
In the case of a complete absence of CME kinematics measurements, Fig. \ref{fig_vr_vs_sxr} is able to reproduce and complement the study of \citet{Salas-Matamoros_2015}. This allows us to obtain, using Eq. \ref{eq_vr_max_sxr} from panel (b), an estimate of the maximum radial speed from the maximum of the associated SXR flare.
We also looked for a more direct relationship between HXR and CME radial acceleration, such as \citet{Berkebile_2012}, but our results were not compelling.\\

We note also that this paper ought to be used as a first step to better understanding global space weather. In fact, when CMEs arrive in the magnetosphere, their interactions with the Earth's magnetic field trigger space weather effects \citep{Pulkkinen_2007}, which are synchronous with the arrival of SEPs. The relations presented could prove very useful for space weather predictions during events that are observed by only one spacecraft or completely missed due to unexpected data gaps in remote-sensing instruments.

\begin{acknowledgements}
    We thank the anonymous reviewer for valuable comments that have improved the manuscript.
    This project has received funding from the European Union’s Horizon 2020 research and innovation program under grant agreement No 101004159 (SERPENTINE project, \url{https://serpentine-h2020.eu/}).
    The SoHO/LASCO CME catalogue is generated and maintained at the CDAW Data Center by NASA and The Catholic University of America in cooperation with the Naval Research Laboratory. SOHO is a project of international cooperation between ESA and NASA.
    The IRAP team acknowledges support from the space weather team in Toulouse (Solar-Terrestrial Observations and Modelling Service, STORMS; \url{http://storms-service.irap.omp.eu/}). This includes funding for the data mining tools AMDA (\url{http://amda.cdpp.eu/}) and the propagation tool (\url{http://propagationtool.cdpp.eu}). A.K. acknowledges financial support from NASA's NNN06AA01C (SO-SIS Phase-E) contract.
\end{acknowledgements}

\bibliographystyle{aa} 
\bibliography{biblio.bib} 

\end{document}